\documentclass[12pt]{article}
\usepackage{amsmath}
\usepackage{amssymb}
\usepackage{comment}
\usepackage[papersize={8.5in,11in}]{geometry}
\allowdisplaybreaks
\renewcommand{\vec}[1]{{\bf{#1}}}

\begin{document}

\title{The virial theorem and planetary atmospheres}

\author{Viktor T. Toth$^\star$\\~\\
{\rm
\footnotesize
$^\star$Ottawa, Ontario K1N 9H5, Canada}}

\maketitle

\begin{abstract}
We derive a version of the virial theorem that is applicable to diatomic planetary atmospheres that are in approximate thermal equilibrium at moderate temperatures and pressures and are sufficiently thin such that the gravitational acceleration can be considered constant. We contrast a pedagogically inclined theoretical presentation with the actual measured properties of air.
\end{abstract}

In his widely discussed article, Miskolczi \cite{Miskolczi2007} postulates that the virial theorem, which relates the average kinetic and average potential energies of a bound mechanical system (see \cite{LL1972} for a thorough introduction), can be applied to a planetary atmosphere in equilibrium in the planet's gravitational field.

To investigate if Miskolczi's postulate is correct (whether or not the postulate was correctly applied in \cite{Miskolczi2007} is a question beyond the scope of the present paper), let us first consider the case of a bouncing ball in an homogeneous gravitational field. (The gravitational field within the Earth's atmosphere, the thickness of which is small compared to the Earth's radius, is approximately homogeneous. However, the same results presented here could also be obtained using a Newtonian gravitational potential \cite{Pacheco2003}\footnote{The author wishes to thank Gy\"orgy Major for bringing Ref.~\cite{Pacheco2003} to his attention.}.)

Consider dropping the ball from a height $h_b$ above the surface, and assume that it falls without air resistance, and bounces back from the ground with no loss of kinetic energy. We wish to calculate its average kinetic and average potential energy. It is sufficient to compute these averages for the first part of the ball's motion, as it falls to the ground; the bounce-back is just a time-reversed copy of its initial drop, and afterwards, in the absence of dissipative losses, the ball repeats the same motion {\it ad infinitum}.

We presume that the ball was dropped at $t=0$. At any other time $t>0$ before the ball hits the ground, its height will be
\begin{equation}
h = h_b-\frac{1}{2}gt^2,
\end{equation}
where $g$ ($\simeq 9.81$~m/s$^2$ on the Earth) is the surface gravitational acceleration. From this,
\begin{equation}
t=\sqrt{\frac{2(h_b-h)}{g}},
\end{equation}
and notably, the time it takes to reach the ground ($h=0$) is
\begin{equation}
t_0=\sqrt{\frac{2h_b}{g}}.
\end{equation}

The velocity of the ball at time $t$ ($0\le t\le t_0$) is
\begin{equation}
v=gt=\sqrt{2g(h_b-h)}.
\end{equation}

The kinetic energy $K$ and potential energy $U$ of the ball are calculated the usual way:
\begin{align}
K=&\frac{1}{2}mv^2=\frac{1}{2}mg^2t^2,\label{eq:K}\\
U=&mgh=mgh_b-\frac{1}{2}mg^2t^2.\label{eq:U}
\end{align}
According to the virial theorem, for a bound mechanical system with kinetic energy $K$,
\begin{equation}
2\langle K\rangle=\left\langle\sum_a \vec{r}_a\cdot\vec{F}_a\right\rangle,\label{eq:VT}
\end{equation}
where $\langle\rangle$ denotes time averaging and $\vec{r}_a$ and $\vec{F}_a$ are the position of, and the force acting on, the $a$-th particle that constitutes the system. In the case of a system in which the potential energy is an homogeneous function of degree $k$ of the coordinates, we get
\begin{equation}
2\langle K\rangle=k\langle U\rangle.
\end{equation}
For (\ref{eq:U}), $k=1$. The time averages of the kinetic energy (\ref{eq:K}) and potential energy (\ref{eq:U}) between $t=0$ and $t=t_0$ can be calculated as
\begin{align}
\langle K\rangle=&\frac{1}{t_0}\int_0^{t_0}K=\frac{1}{6}mg^2t_0^2=\frac{1}{3}mgh_b,\\
\langle U\rangle=&\frac{1}{t_0}\int_0^{t_0}U=mgh_b-\frac{1}{3}mgh_b=\frac{2}{3}mgh_b.
\end{align}
Hence, $2\langle K\rangle=\langle U\rangle$ and the virial theorem for the potential (\ref{eq:U}) appears satisfied. This was made possible, in part, by referencing the height $h$ to ground level; this allowed us to ignore the effects of the ground surface in (\ref{eq:VT}), because at the ground, $\vec{r}_a=0$, as observed also in \cite{Pacheco2003}.

How about a column of atmospheric gas? We assume a column of gas in hydrostatic equilibrium standing over a unit surface area in an homogeneous gravitational field. Its density will be a function of height $h$ above ground:
\begin{equation}
\rho = \rho(h).
\end{equation}
The pressure at $h$ is equal to the weight of gas situated at heights above $h$:
\begin{equation}
p(h)=\int_h^\infty g\rho(h')~dh'.\label{eq:prho}
\end{equation}

We assume that the gas is in thermal equilibrium (the real atmosphere isn't, but that's another story), so its temperature is constant:
\begin{equation}
T=T_0.
\end{equation}
We also assume that the gas obeys the ideal gas law (this is a valid approximation for air at room temperature and sea level pressure), hence
\begin{equation}
pV = nRT,
\end{equation}
where $V$ is the volume of $n$ moles of gas, and $R\simeq 8.31$~JK$^{-1}$mol$^{-1}$ is the ideal gas constant. The mass of $n$ moles of gas is $nM_n$ where $M_n$ ($\simeq 0.029$~kg/mol for air) is the molar mass of the gas; its density is $\rho=nM_n/V$, hence $V=nM_n/\rho$. We can thus rewrite the ideal gas law in the form
\begin{equation}
p=\frac{RT_0}{M_n}\rho.
\end{equation}
Using this in (\ref{eq:prho}), we obtain
\begin{equation}
\frac{RT_0}{M_n}\rho=\int_0^\infty g\rho(h')~dh'-\int_0^hg\rho(h')~dh',
\end{equation}
or, in differential form,
\begin{equation}
\frac{RT_0}{gM_n}\frac{d\rho}{dh}=-\rho,
\end{equation}
which can be solved trivially:
\begin{align}
-\frac{RT_0}{gM_n}\frac{1}{\rho}d\rho=dh,\\
-\frac{RT_0}{gM_n}\log\rho=h+C,
\end{align}
or
\begin{equation}
\rho=\rho_0e^{-gM_nh/RT_0},
\end{equation}
which we can also write in the form
\begin{align}
\rho=&\rho_0e^{-h/h_0},\\
h_0=&\frac{RT_0}{gM_n}.\label{eq:h0}
\end{align}
For air at $T_0=273$~K, we get
\begin{equation}
h_0=\frac{8.31\times 273}{9.81\times 0.029}\simeq 8~\mathrm{km},
\end{equation}
which agrees well with the observed properties of the atmosphere.

So what about the virial theorem? Going back to the bouncing ball for a moment, we can immediately spot a potential problem: what if the ball is moving horizontally as well? Indeed, it can move horizontally at an arbitrary velocity, yet its potential energy will be no different, hence the virial theorem fails. We must make sure that we only consider the vertical component of the ball's velocity before the theorem can be considered valid. The velocity of the ball can be written in rectilinear form as $v^2=v_x^2+v_y^2+v_z^2$, but we are only interested in the vertical component. In the specific case when the average velocities in the $x$, $y$, and $z$ direction are the same, we get $\langle v_z^2\rangle=\frac{1}{3}\langle v^2\rangle$. Accordingly, the virial theorem in this case reads
\begin{equation}
\frac{2}{3}\langle K\rangle=\langle U\rangle.
\end{equation}

This result can also be obtained using another argument, presented in \cite{Pacheco2003}: rather than allowing the coordinates to remain unbounded in the horizontal plane, we can consider confining the ball to within a cylinder of unit radius, integrating and time averaging the force acting on the ball as it hits the cylinder walls, in order to obtain the right-hand side of (\ref{eq:VT}).

An atmosphere, unfortunately, is not made of bouncing balls, however appealing that picture might appear. Air, in particular, is composed mainly of diatomic gases (notably N$_2$ and O$_2$), which at room temperature have two rotational degrees of freedom in addition to the three translational degrees of freedom discussed above. (At higher temperatures, vibrational modes also play a role.) The kinetic energy of a column of gas is its internal thermal energy. The principle of equipartition of energy states that internal thermal energy is distributed equally between all degrees of freedom. Therefore, the virial theorem now reads
\begin{equation}
\frac{2}{5}{\langle K\rangle}={\langle U\rangle}.\label{eq:virialair}
\end{equation}
This is our main result, valid for any diatomic atmosphere that obeys the ideal gas law in an homogeneous gravitational field at moderate temperatures.

For a column of gas over a unit surface area, the thermal kinetic energy is
\begin{equation}
K=\int_0^\infty c_VT_0\rho~dh=c_V\rho_0T_0\int_0^\infty e^{-h/h_0}~dh=c_V\rho_0T_0h_0,\label{eq:Kgas}
\end{equation}
where $c_V$ is the specific heat of the atmosphere at constant volume. The potential energy is just the gravitational potential energy:
\begin{equation}
U=\int_0^\infty g\rho h~dh=g\rho_0\int_0^\infty he^{-h/h_0}~dh=g\rho_0h_0^2.\label{eq:Ugas}
\end{equation}
The ratio of the two is
\begin{equation}
\frac{U}{K}=\frac{gh_0}{c_VT_0}.
\end{equation}
Using (\ref{eq:h0}), we obtain
\begin{equation}
\frac{U}{K}=\frac{R}{c_VM_n}.\label{eq:UK}
\end{equation}
Given (\ref{eq:virialair}) and (\ref{eq:UK}), we can calculate the specific heat $c_V$. For air, we obtain
\begin{equation}
c_V=\frac{5R}{2M_n}\simeq 716~\frac{\mathrm{J}}{\mathrm{K}\cdot\mathrm{kg}}.
\end{equation}
a value that agrees well with the known properties of air ($c_V=718$~JK$^{-1}$kg$^{-1}$).

In this derivation, we assumed that $T=T_0$ is constant. Our result, however, remains valid even when $T$ is not constant, so long as the gas is in ``local thermodynamic equilibrium'', which ensures that the principle of equipartition remains valid and that thermodynamic quantities, such as the temperature or specific heat, remain well-defined. To see this, we first rewrite the condition of hydrostatic equilibrium (\ref{eq:prho}) in differential form:
\begin{equation}
dp=-g\rho~dh.
\end{equation}
This allows us to write the potential energy of the gas (\ref{eq:Ugas}) as
\begin{equation}
U=\int_0^\infty g\rho h~dh=-\int_0^\infty h~dp=-\int_0^\infty h\frac{dp}{dh}~dh=\int_0^\infty p~dh,\label{eq:Up}
\end{equation}
where the last step was taken by integrating in parts and using $p(\infty)=0$. On the other hand, the thermal kinetic energy (\ref{eq:Kgas}) can be rewritten as
\begin{equation}
K=c_V\int_0^\infty T\rho~dh=\frac{c_VM_n}{R}\int_0^\infty p~dh.\label{eq:Kp}
\end{equation}
The ratio of (\ref{eq:Up}) and (\ref{eq:Kp}) remains the same constant ratio (\ref{eq:UK}) that we obtained in the $T=T_0$ case:
\begin{equation}
\frac{U}{K}=\frac{R}{c_VM_n},
\end{equation}
even as $T$ varies with altitude. Therefore, even as we allow $T$ to be a function of $h$, (\ref{eq:virialair}) remains satisfied.

Hence we were able to demonstrate without having to invoke concepts such as ``hard core'' potentials or intramolecular forces that the virial theorem is indeed applicable to the case of an atmosphere in hydrostatic equilibrium. However, it must be ``handled with care'': the nature of the atmosphere and the fact that the horizontal (translational) and internal (rotational) degrees of freedom of the gas molecules are unrelated to the gravitational potential cannot be ignored.

\bibliography{refs}
\bibliographystyle{unsrt}

\end{document}